\documentclass[twocolumn,showpacs,preprintnumbers,amsmath,amssymb,superscriptaddress]{revtex4}
\usepackage{mathrsfs}
\usepackage{graphicx}
\usepackage{subfigure}
\usepackage{dcolumn}
\usepackage{bm}
\usepackage{amsmath}
\usepackage{mathtools}
\usepackage{xfrac}

\begin{document}

\title[Aviram-Ratner DNA base-pair rectifiers]
{Aviram-Ratner rectifying mechanism for DNA base-pair sequencing
through graphene nanogaps}
\author{Luis A. Agapito} \affiliation {Department of Physics and W. M. Keck Computational Materials Theory Center,
California State University Northridge, Northridge, CA 91330, USA}
\email{luis.agapito@gmail.com}
\email{nick.kioussis@csun.edu}
\author{Jacob Gayles} \affiliation {Department of Physics, Texas A\&M University, College Station, TX 77843, USA}
\author{Christian Wolowiec} \affiliation {Department of Physics, University of California San Diego, La Jolla, CA 92093, USA}
\author{Nicholas Kioussis} \affiliation {Department of Physics and W. M. Keck Computational Materials Theory Center, California State University Northridge, Northridge, CA 91330, USA}


\begin{abstract}
We demonstrate that biological molecules such as Watson-Crick DNA base pairs can behave as biological Aviram-Ratner electrical rectifiers because of the spatial separation and weak hydrogen bonding between the nucleobases. We have performed a parallel computational implementation of the \textit{ab-initio} non-equilibrium Green's function (NEGF) theory to determine the electrical response of graphene---base-pair---graphene junctions. The results show an asymmetric (rectifying) current-voltage response for the Cytosine-Guanine base pair adsorbed on a graphene nanogap. In sharp contrast we find a symmetric response
for the Thymine-Adenine case. We propose applying the asymmetry of the current-voltage response as a sensing criterion to the technological challenge of rapid DNA sequencing via graphene nanogaps.
\end{abstract}


\maketitle

\section{Introduction}
Rapid and low-cost genome sequencing is one of the grand challenges of genome science today. The Sanger
method \cite{sanger} has served as the cornerstone for genome sequence production since 1977, close to almost
30 years of tremendous utility. With the
completion of the human genome
sequence \cite{venter,lander}, there is an
imminent need in developing new sequencing methodologies that will enable
``personal genomics" or the routine study of our individual genomes \cite{chan,zwolak2008}.

One potential candidate is nanopore sequencing \cite{kasianowicz,
akeson, deamer}, where a negatively-charged single-stranded (ss) DNA
(in solution with counterions) is envisioned to translocate through
$\alpha$-hemolysin channels in lipid bilayers \cite{kasianowicz,
akeson, deamer} or through solid-state nanopores \cite{li, storm,
harrell, lemay, biance, zwolak2005}, by a longitudinal electric field
from one side of a membrane to the other. As the nucleotides of the
DNA are migrated across the membrane, a significant fraction of ions
are blocked from simultaneously entering the pore depending on their
size. By continuously measuring the ionic blockade current, single
molecules of DNA may be detected. Another proposed electronic
approach is based on the intrinsic electronic properties of the
bases by, for instance, embedding electrodes within a nanopore to
measure the transverse current through ssDNA as it
translocates through the pore \cite{zwolak2005}. However, the
construction of a nanopore with embedded electrodes remains a
formidable challenge to the implementation and testing of this
method.

Recently, a systematic nanoelectrode-gated transverse
electron-tunneling molecular detection concept with potential for
rapid DNA sequencing has been proposed \cite{lee07}. According to
the nanoelectrode-gated molecular-detection concept, it should be
possible to obtain genetic sequence information by probing a DNA
molecule base by base at a sub-nanometer scale. The nanoscale
reading of DNA sequences is envisioned to take place at a nanogap
defined by a pair of nanoelectrode tips as a DNA molecule moves
through the gap base by base. The rationale is that the four
different nucleotide bases [adenine (A), thymine (T), guanine (G),
and cytosine (C)] and their various sequences, each with a distinct
chemical composition and structure, should be associated with a
specific signature of transverse tunneling current across the two
tips. Theoretically, this new approach has the potential to sequence
DNA at a maximal rate of 106 base pairs per second per detection
gate. This method can be extended to parallel arrays of
multiple nanoelectrode detection gates, thus magnifying the readout
throughput by additional orders of magnitude, achieving estimated
maximal rates of possibly hundred million (100 Mb) bases per second
per device. Because the nanoelectrode-gate electron tunneling
detection approach does not depend on the transient blockade of the
ion current, one can use a wider range of detection gap sizes (1.5 -
5 nm) compared to the very small detection gap sizes (1.5 - 2.5 nm)
required by the electrophysiology-like approaches.

Despite the promise that transverse conduction measurements of DNA
molecules holds for rapid sequencing, a comprehensive experimental
study of transverse DNA conductance showing a variation of the
conductance with base type has not been performed yet. Furthermore, the exact
mechanism of electronic transport and its signature
is debated \cite{zwolak2005, zikic1, lagerqvist, lagerqvist2}.

In this manuscript we study a novel approach for transverse charge
transport through individual or a sequence of bases of double
stranded (ds) DNA placed between the fringes of a graphene
nanogap. The electronic signatures of the individual base or
sequence of bases will display characteristic discrete peaks in the
density of states, which will lead to a varying conductance
fingerprint as the bias voltage is varied.  The novel single-layer
graphene-nanogap technique proposed here for transverse conductance measurements
of molecules offers the following desirable features compared to
commonly-used techniques employing thin metallic (Au) wires: (1)
accurate imaging of the location of the molecules inside the
nanogap; (2) precise knowledge of the configuration of all atoms in
contact with the molecule; (3) planar geometry which allows for the
fabrication of parallel arrays of graphene nanogaps, thus magnifying the DNA
sequencing rate; and (4) tuning of the discrete molecular energy
levels with respect to the leads' chemical potential.

We demonstrate that the DNA base pair behaves as a biological Aviram-Ratner electrical rectifiers because of the spatial separation and weak bonding between the nucleobases.

\section{Material models}
\subsection{The two-level model}
In 1974 Aviram and Ratner \cite{Aviram1974} showed that a molecular
dimer composed of both an acceptor and a donor subunits that are
spatially separated by a sigma-bond barrier behave as a molecular
rectifier. The two-level model in Figure 1a shows the relative
position of the highest occupied HOMO (\textbf{2}, \textbf{4}) and
lowest unoccupied LUMO molecular orbitals (\textbf{1}, \textbf{3}),
with respect to the Fermi level $E_F$, for the acceptor and
donor subunits of a generic Aviram-Ratner system. The application of
a bias voltage $V$ across the junction causes a rigid
shift of these energy levels by $\mp\frac{V}{2}$ depending on
whether they are spatially localized on the left or right side of
the junction, as seen in Figure 1b. The relative shift of the levels
is responsible for the rectifying behavior, which is characterized
by an asymmetric current-voltage ($I-V$) curve shown in Figure 1c.

For forward bias, the current rises around $V_F =
\frac{1}{e}\min\{\Delta_{\mathbf{1}},\Delta_{\mathbf{4}}\}$ as
either \textbf{1} or \textbf{4} enters to the Fermi energy window, highlighted in magenta. The schematic Figure 1b shows a tunneling processes that is resonant through the donor level \textbf{4} but
non-resonant across the acceptor subunit. For reverse bias, analogously, the current onset happens at $V_R =
\frac{1}{e}\min\{\Delta_{\mathbf{2}},\Delta_{\mathbf{3}}\}$. The
donor-acceptor condition implies that the energies of the frontier levels of the subunits satisfy $\varepsilon_{\mathbf{2}} <
\varepsilon_{\mathbf{4}} < \varepsilon_{\mathbf{1}} <
\varepsilon_{\mathbf{3}}$; consequently $V_F < V_R$, which causes the typical $I-V$ asymmetry of Aviram-Ratner diodes.

\subsection{The single-base-pair model for transverse tunneling}
\subsubsection{Longitudinal hopping vs. transverse tunneling}
The  longitudinal electronic transport along DNA remains controversial and has not been conclusively determined whether the DNA is metallic or insulating \cite{Endres2004,Taniguchi2006}. Longitudinal transport directly involves the fluctuating chemical environment around the DNA backbones and thus can support multiple charge transfer mechanisms that arise from the small activation gaps induced by water and counterions.

In contrast, the transverse electronic transport, perpendicular to the dsDNA axis, is a simpler and less controversial process. It involves an insulating barrier (the hydrophobic core) and only a few discrete energy levels within the barrier, which belong to the base pair.

\subsubsection{The single-base-pair approximation}
The interaction between the stacked base pairs is negligible (of the order of 0.01 eV for A-DNA and 0.1 eV for B-DNA \cite{Endres2002,Endres2004} and therefore the total transverse current of DNA translocating through a nanopore can be well approximated as independent contributions from multiple channels. Each base pair temporarily located within the nanopore's electrodes (recognition region) constitutes an independent channel. In the case of zero-thickness graphene electrodes, our model approximates transverse tunneling as through a single base pair that is decoupled from its neighbors, as shown in Figure 2. The ionic environment around the backbone (ions, counterions, solvent) and the dynamics of the translocating process are necessary for a complete description of in vivo DNA; nonetheless, they involve computationally intense calculations which are out of reach currently employing solely \textit{ab initio} calculations. In addition, as discussed later, both experiment and theoretical calculations have shown that the transverse transport primarily depends on the nature of the nucleobases rather than on the \textit{in vivo} environment. Our proof-of-concept model attempts to address the underlying physics of transverse transport and focuses on single base pairs.

\subsubsection{Effect of the backbone, solvent, and counterions}
\textit{Stability of the energy levels of the base pairs:}\\
Regarding the solvent, fluctuations of the surrounding water molecules are known to have little effect on the transverse current for the case of ssDNA, amounting to small modulation of its magnitude \cite{lagerqvist2}. More importantly, the energy levels for dsDNA are expected to be more stable against external perturbations compared to those of the ssDNA case, as each base pair is protected inside the hydrophobic core that is further stabilized by the interaction between backbones and counterions. In cases when water can enter into the DNA structure through damaged sites, it can induce small activation gaps around the Fermi level \cite{Hubsch2005}.

In regard to the backbone, the discrete energy levels of the base-pairs, relevant to transport, are electronically isolated from the continuous spectrum of the backbone's phosphate and sugar (i.e. off-resonance \cite{Krems2009}), which in turn have a much larger energy gap of $\sim$ 6.5 eV \cite{Jauregui2009} and hence do not hybridize significantly with the base pair within the relevant energy window \cite{Min2011,Xu2005}. The presence of counterions (e.g. Na$^+$, Mg$^+$) is more critical, however, since they introduce energy levels in the base-pair's HOMO-LUMO gap by hole doping \cite{Sharpir2007,Endres2004}. Nonetheless, these states are localized outside the backbones and could induce electron hopping (in conjunction with nearby water) along the longitudinal direction but are unlikely to contribute to transverse tunneling, which proceeds primarily through the hydrogen-bonded base pairs. Recent experiments find that the leakage current due to these localized states amounts to a small offset and noise in the measured tunneling current and is readily separable from the main contribution coming from tunneling through the nucleobases \cite{Tsutsui2010}.

Furthermore, shift of the discrete levels due to geometrical deviations from the optimal adsorption ($\pi$-stacking) of the base pair on the graphene electrodes caused by thermal fluctuations or backbone-strain relaxation (not considering translocation) is also negligible. Combined molecular dynamics (MD) and density functional theory (DFT) statistical analysis of nucleobase adsorption on graphene \cite{Cho2011,Min2011} and CNT \cite{Meng2006} confirm that this type of perturbation has small effect and that the relative positions of the nucleobase's levels are locked, almost independently from the orientation between the nucleobases and the electrode.

\textit{Alignment between the energy levels and the electrode's Fermi level:}\\
Another relevant factor in the Aviram-Ratner mechanism, besides the stability of the discrete energy levels discussed above, is the alignment between these levels and the Fermi level of the electrodes $E_F$. Such alignment determines the parameters $\Delta_{\mathbf{\it i}}$ ($i$=1-4) and is expected to be susceptible to the relative geometrical orientation between the hydrophobic core and the electrodes. At the equilibrium optimal adsorption, the alignment is mainly determined by a van der Waals interaction. However, far from equilibrium the alignment is generally determined by the long-range Coulomb interaction that couples the nucleobase to the electrodes. The dynamical displacements during DNA translocation creates a random electrostatic environment that overlaps with the weaker nucleobase-electrode interaction which in turn can cause dynamical fluctuations in the HOMO-EF alignment and thus in $\Delta_{\mathbf{\it i}}$. Since the nucleobase-electrode interaction is ultimately dependent on the electrical polarizability of each individual nucleobases (G $>$ A $>$ T $>$ C) \cite{Gowtham2007,Riahi2010,Antony2008}, the values of $\Delta_{\mathbf{\it i}}$
should also follow a consistent trend as long as the polarizability of the nucleobases can be consistently detected in the midst of the environmental electronic noise. Although apparently counterintuitive, MD simulations have proven that such weak nucleobase-electrode interaction can be distinguished from the chaotic environmental electronic noise \cite{Sigalov2007}. Statistical averaging or increase of residence time of the nucleotide inside the recognition region is needed to manage the electronic noise \cite{Huang2010}. Distinguishing such a weak interaction is equivalent to measuring the capacitance of each nucleobase, which is another promising technique for fast DNA sequencing with the single-base resolution \cite{Lu2008}.

\textit{Experimental reproducible determination of the energy levels}:\\
The HOMO-$E_F$ alignment is primarily pinned by the intrinsic electronic signature of each nucleobase. Despite the expected random electrostatic environment of \textit{in vivo} DNA, electrostatic cancelation effects are expected \cite{Min2011}, resulting in a more electrostatically homogeneous medium, and consequently a less fluctuating HOMO-$E_F$ alignment. In that regard, by measuring the transverse-tunneling current (with scanning tunneling spectroscopy STS), experiments have successfully pinpointed the value of the HOMO level with respect to $E_F$ within a 10\% and 4\% standard deviation for poly(CG) and poly(TA) dsDNA, respectively \cite{Xu2007,Xu2007-2}.

Moreover, the same STS technique has also reproducibly measured the nucleotide's HOMO levels in ssDNA \cite{Tanaka2009,Huang2010}, despite the fact that ssDNA is conformational less stable than dsDNA. Furthermore, a more recent experiment using a simple break junction instead of a STM has further proved that such identification is possible, a step closer to our proposed junction model \cite{Tsutsui2011}.

In short, recent STS measurements have shown that it is possible to pin the energy levels of nucleobases with encouraging degree of reproducibility, despite the presence of backbone, solvent, and counterions. Moreover, an Aviram-Ratner rectifying shape of the transverse current, similar to that predicted in this manuscript, has already been experimentally observed in poly(CG) dsDNA \cite{Sharpir2007}.

\section{Computational methods}
All the electronic structure calculations are performed at the {\it ab initio} level using the SIESTA method \cite{Soler2002} with
the generalized-gradient Perdew-Burke-Ernzerhof exchange-correlation density functional \cite{perdew}. The valence electrons are expanded using
localized basis of double-$\zeta$ with polarization orbitals (DZP). We found that polarization orbitals are especially needed to screen the applied bias voltage along the electrodes.
Troullier-Martin norm-conserving pseudopotentials \cite{Troullier1991} in the Kleinman-Bylander form \cite{Kleinman1982} are used for the core electrons.

The electronic density is considered converged after the maximum difference between the density matrices of two consecutive cycles is smaller than $10^{-5}$.

The positions of the atoms in the unit cell of the nanoribbon (a = 4.92\AA), the reconstruction of the nanogap, and the adsorption geometry of the base-pairs on the nanogap were self-consistently optimized by conjugate gradient until the forces are less than 0.01 eV/\AA.

For the isolated base pairs, ample vacuum separation was included inside the unit cells to avoid interaction with mirror images.  In the case of the infinite nanoribbon, the calculations used 80 k-points along the periodic direction. For the case of the finite C-G and T-A junctions, in order to speed up the convergence of the wavefunctions, the boundary atoms in the unit cell are made to match periodically along the longitudinal direction; ample vacuum is included in the other two directions.

\subsection{Non-equilibrium Green's function (NEGF)}
The central part of the NEGF method involves calculating the electron density of an infinite junction under bias voltage. This subsection presents a self-contained derivation leading into a matrix representation of the non-equilibrium electron density, given in Eq. \ref{dm_app}. A matrix representation is important since all necessary quantum-mechanical quantities such as Hamiltonian and overlap operators can be readily extracted from any localized-orbital DFT package in matrix form.

The spectral-function operator $\hat{A}$ of a system with Hamiltonian operator $\hat{H}$ is defined as

\begin{equation}
 \hat{A}(\varepsilon)\equiv 2\pi\delta(\varepsilon - \hat{H})\label{sf}
\end{equation}

From the Sokhotsky-Weierstrass theorem, the retarded (+) and advanced (-) Green's function operators $\hat{G}^{\pm}(\varepsilon)$ can be decomposed as:

\begin{equation}
 \hat{G}^{\pm}(\varepsilon)\equiv \lim_{\eta \to 0} \frac{1}{\varepsilon -\hat{H} \pm \eta i} = \mathcal{P}\frac{1}{\varepsilon -\hat{H}} \mp i\pi\delta(\varepsilon - \hat{H}),
\end{equation}
where $\mathcal{P}$ represents the Cauchy principal-value of an associated integral. Using this relation, the spectral-function operator reduces to

\begin{equation}
 \hat{A}(\varepsilon)\equiv 2\pi\delta(\varepsilon - \hat{H}) = i(\hat{G}^{+}-\hat{G}^{-})\label{spec_op}
\end{equation}

The density-matrix operator $\hat{D}$ is defined as the Fermi function $f(\hat{H}-\mu)=[\exp{((\hat{H}-\mu)/(k_BT))]}+1]^{-1}$ of the Hamiltonian operator $\hat{H}$ at the ``equilibrium'' chemical potential $\mu$

\begin{align}
\hat{D}(\mu) &\equiv f(\hat{H}-\mu)\\
&=\int_{-\infty}^{+\infty} f(\varepsilon - \mu)\delta(\varepsilon - \hat{H})d\varepsilon\\
\bar{D}(\mu) &=\frac{1}{2\pi}\int_{-\infty}^{+\infty} d\varepsilon\,f(\varepsilon - \mu)\bar{A}(\varepsilon)\label{dm0}
\end{align}

The bars in the last expression denote matrices expanded in an orthogonal basis set. Otherwise, all matrices are given non-orthogonal basis by default.

For the case of an ``infinite'' junction, the retarded Green's function is obtained by including the effect of the left ($L$) and right ($R$) semi-infinite electrodes through self-energy corrections $\Sigma_{\{L,R\}}$  to the the finite system $H$ \cite{Agapito2007}, resulting in the following expression

\begin{equation}
\bar{G}^+ = G^+S = [\varepsilon^{+}S- H -\Sigma_L(\varepsilon^{+}) -\Sigma_R(\varepsilon^{+})]^{-1}S
\end{equation}

This leads to the following expression

\begin{align}
\begin{split}
{(G^{+})}^{-1}-(G^{-})^{-1} &= \varepsilon^{+}S-H-\Sigma_L(\varepsilon^+)-\Sigma_R(\varepsilon^+)\\
&-(\varepsilon^{-}S-H-\Sigma_L(\varepsilon^-)-\Sigma_R(\varepsilon^-))\\
&=(\varepsilon^+ - \varepsilon^-)S+i(\Gamma_L+\Gamma_R)\\
\intertext{where the definition of the broadening matrix $\Gamma_{\{L,R\}}\equiv i(\Sigma_{\{L,R\}}(\varepsilon^+)-\Sigma_{\{L,R\}}(\varepsilon^-))$ was used in the last line . Multiplying both sides of the equation through the left and right by $-iG^+$ and $G^-$, respectively, one finds}
i(G^+-G^-)&=2\eta G^{+}SG^- + G^+(\Gamma_L+\Gamma_R)G^-\label{gmg}
\end{split}
\end{align}

The Green's function matrix in an orthogonal basis is obtained through $\bar{G}=GS$ \cite{Agapito2008}. Combining Eq. \ref{spec_op} and Eq. \ref{gmg}, one finds

\begin{subequations}
\begin{align}
\bar{A}&=i({\bar{G}}^{+}-\bar{G}^-)\label{sf_a}\\
\bar{A}&=2\eta G^{+}SG^-S + G^+(\Gamma_L+\Gamma_R)G^-S\label{sf_b}
\end{align}
\end{subequations}

By definition, the density of states of the junction per spin channel is given by
\begin{subequations}
\begin{align}
DOS \equiv Tr(\frac{A}{2\pi})=Tr(\frac{\bar{A}S^{-1}}{2\pi})
\end{align}
\end{subequations}

In the absence of coupling of the central region to the electrodes ($\Gamma_L=\Gamma_R=0$), it follows from Eq. \ref{sf_b} that the density of states of the junction reduces to \cite{Li2007}

\begin{equation}
DOS\big|_{\Gamma_{\{L,R\}}=0}=Tr(\sum_{n}c_n \frac{1}{\pi}\frac{\eta}{(\varepsilon-\varepsilon_n)^2+\eta^2}\phi_n\phi_n^{\dagger})\label{dos}
\end{equation}
\\where $c_n=\phi_n^{\dagger}S\phi_n$. $\varepsilon_n$ and $\phi_n$ are the generalized eigenvalues and eigenvectors of the finite system defined by $H$ and $S$. From Eq. \ref{dos}, it can be verified that the density of states of a uncoupled junction reduces to a series of discrete energy levels represented by Lorentzian distributions with $\eta\to0$.

In the presence of coupling, the energy levels of the central region are no longer discrete but continuous in energy, as schematically depicted by the broad horizontal lines in Figure \ref{fig_complex_plane}a. From Eq. \ref{sf_b}, the ``scattered'' electronic states of the infinite junction fall into one of the following cases: (\textit{i}) states that couple the central region and left lead (the term $G^+\Gamma_LG^-S$); (\textit{ii}) the central region and the right lead (the term $G^+\Gamma_RG^-S$); the states that couple the central region and both leads are projected into the previous two cases; (\textit{iii}) states localized in the central region that do not couple to any of the leads (the term $2\eta G^{+}SG^-S$); and (\textit{iv}) states that couple both leads but not the central region, which are noticeably absent in Eq. \ref{sf_b} as a consequence of the underlying assumption of non-iteracting electrodes.

The density matrix gives the electronic charge of the junction and is found by filling (integrating) the scattered states over the energy range where reservoir electrons are available in the electrodes, as expressed by Eq. \ref{dm0}. Without applied bias voltage, the states corresponding to cases \textit{i} and \textit{ii} are evenly filled up to the chemical potential of the junction ($\mu=E_F$).

Under applied $V$, however, states \textit{i} and \textit{ii} are unevenly filled since they are required to reach local equilibrium with the chemical potential of the left ($\mu_{L}=E_F-\frac{V}{2}$) or right ($\mu_R=E_F+\frac{V}{2}$) electrode, respectively. As schematically shown in Figure \ref{fig_complex_plane}a, the total electronic charge of the central region derives from the states in the green and gray areas . Therefore, the charge is said to be decomposed into an ``equilibrium'' (green) and a ``non-equilibrium'' (gray) contribution.

The equilibrium contribution $\bar{D}_{eq}$ is analogous to a fictitious zero-bias problem (given by Eq. \ref{dm0}) where both electrodes have the same chemical potential $\mu_L$. Then, $\bar{D}_{eq}=\bar{D}(\mu_L)$. In this fictitious zero-bias problem some information of the applied bias voltage is implicitly included in the shifted electronic structure of the right electrode. The non-equilibrium contribution $\bar{D}_{non-eq}$ results from all the states inside the gray area, which do not couple to the left electrode and therefore can be obtained from the spectral function in Eq. \ref{sf_b} when $\Gamma_L = 0$. Following Eq. \ref{dm0}, we compute this contribution by integrating $\bar{A}\big|_{\Gamma_L=0}$ over the $f(\varepsilon - \mu_R)-f(\varepsilon - \mu_L)$ energy window. Combining Eq. \ref{dm0} and Eq. \ref{sf_b}, the
density matrix for the ``infinite'' junction under bias voltage is:

\begin{align}
\begin{split}
\bar{D}(\mu_L,\mu_R) &= \bar{D}_{eq} + \bar{D}_{non-eq}\\
&=\frac{1}{2\pi}\int\limits_{-\infty}^{+\infty} d\varepsilon\,i({\bar{G}}^{+}-\bar{G}^-)f(\varepsilon - \mu_L)\\
&+\frac{1}{2\pi}\int\limits_{-\infty}^{+\infty} d\varepsilon\,[2\eta G^{+}SG^-S + G^+\Gamma_RG^-S]\\
&\times[f(\varepsilon - \mu_R)-f(\varepsilon - \mu_L)]\label{dm_app}
\end{split}
\end{align}

\begin{figure}
\includegraphics[width=.4\textwidth]{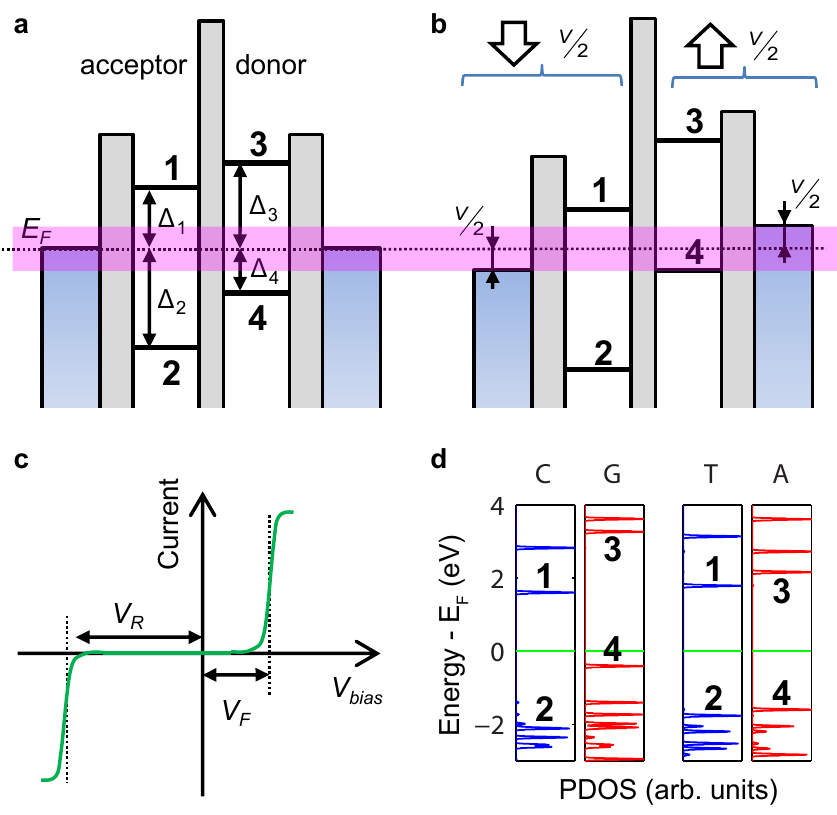}
\caption{\label{fig1} Two-level model of an Aviram-Ratner molecular
rectifier. (a) \textbf{2} and \textbf{1} (\textbf{4} and \textbf{3})
are the HOMO and LUMO levels of the acceptor (donor) subunit. (b)
Rigid shift of the energy levels under an external bias voltage
$V=\frac{1}{e}\Delta_{\mathbf{4}}$ applied across the electrode's reservoir states (shown in blue). Both, the
interfacial and the internal energy barriers are shown in gray. The Fermi energy window due to $V$ is shown in magenta. (c) Asymmetric current-voltage curve of an Aviram-Ratner
diode; the forward ($V_F$) and reverse ($V_R$) threshold
voltages for the onset of the current depend directly on the relative
position of the frontier energy levels. (d) \textit{Ab-initio} energy levels
for the isolated C-G and T-A base pairs, obtained from the projection of the
density of states (PDOS) on the pyrimidine (blue) and purine (red)
subunits.}
\end{figure}

\section{Results and discussion}

\subsection{Isolated Cytosine-Guanine (C-G) and Thymine-Adenine (T-A) base pairs}
The original Aviram-Ratner rectifier employs a
sigma-bond internal barrier between the $\pi$-systems of the subunits. For the case of DNA base pairs, both subunits
are weakly coupled by hydrogen bonds that act as internal
insulating barriers. Figure 1d shows the density of states of
isolated C-G and T-A molecules projected on their subunits. The isolated molecules are calculated without electrodes but using the geometry they adopt in the junction environment.

First, the presence of sharp peaks indicates minimal hybridization between the subunits, which is a direct consequence of the weak hydrogen-bond coupling. Therefore, the electronic structure of the base pair can
be approximately viewed as the superposition of the two subunits, which fulfills the barrier condition. Second, the ordering of the computed energy levels ($\varepsilon_{\mathbf{2}} < \varepsilon_{\mathbf{4}} < \varepsilon_{\mathbf{1}} < \varepsilon_{\mathbf{3}}$) corresponds to that of an acceptor-donor pair.
Therefore, both C-G and T-A base pairs are biological Aviram-Ratner
rectifiers. Our \textit{ab-initio} values of $\Delta_{\mathbf{\it i}}$
($i$=1-4) for the isolated C-G and A-T are 1.61 eV, 2.11 eV, 3.27 eV,
0.41 eV, and 1.79 eV, 1.76 eV, 2.16 eV, 1.6 eV, respectively.
For both base pairs $\Delta_{\mathbf{4}} < \Delta_{\mathbf{1}}$ and
$\Delta_{\mathbf{2}} < \Delta_{\mathbf{3}}$. Consequently, the
forward ($V_F = \frac{1}{e}\Delta_{\mathbf{4}}$) and reverse ($V_R =
\frac{1}{e}\Delta_{\mathbf{2}}$) threshold biases are determined
primarily by the HOMOs of the purine (level \textbf{4}) and
pyrimidine (level \textbf{2}) subunits, respectively. Moreover,
because of the large difference between the C and G
HOMO energies ($\Delta_{\mathbf{2}} \gg \Delta_{\mathbf{4}}$), C-G
is expected to exhibit a noticeably more asymmetric $I-V$ curve
($\frac{V_F}{V_R} = 0.19$) than T-A ($\frac{ V_F}{ V_R} = 0.91$). We
propose that this difference in $I-V$ asymmetries can be used as an
electrical probe for discerning between C-G and T-A pair.

\begin{figure}
\includegraphics[width=.4\textwidth]{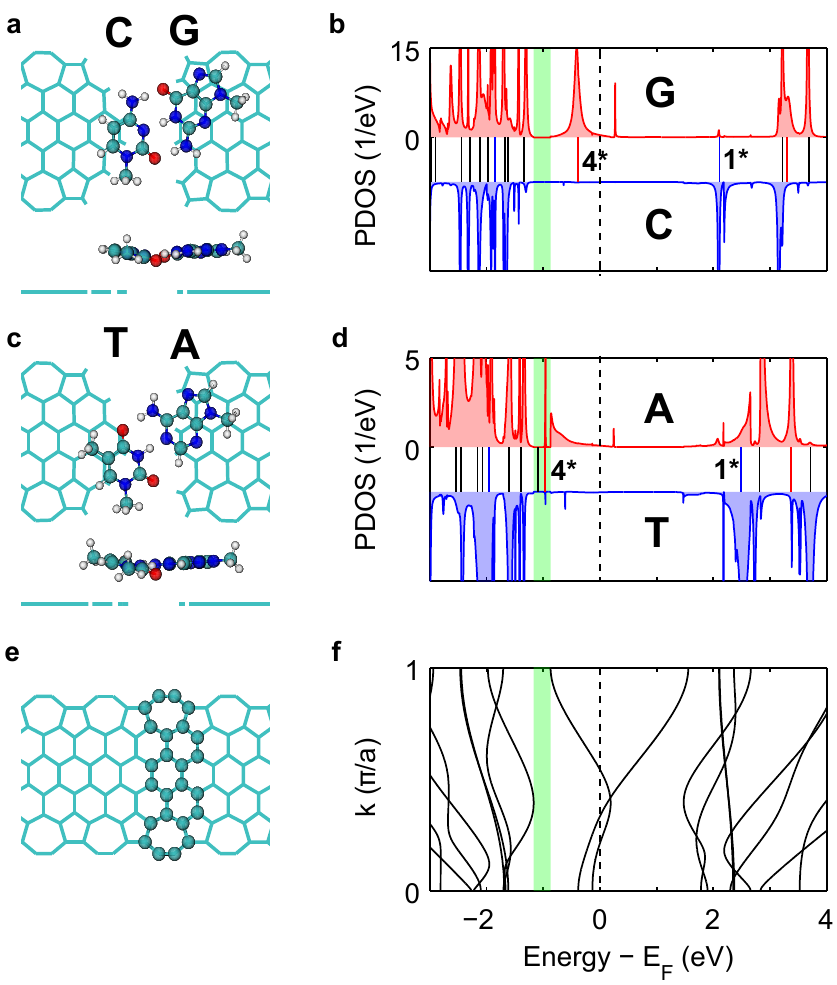}
\caption{\label{fig_bp} DNA base-pair junctions. Top and side view
of the C-G (a) and T-A (c) junctions. The density of states are
projected on the purine (red) and pyrimidine (blue) components of
the C-G (b) and T-A (d) junctions. The projected eigenstates are
shown for reference on the middle panels of (b) and (d); the colored
levels correspond to the four relevant ones in the two-level model.
The relevant projected eigenstate \textbf{4*} is located at $-0.38
(-0.92)$ eV for the C-G (T-A) junction.  The unit cell and
bandstructure for the reconstructed ribbon are shown in (e) and (f);
the ribbon exhibits a forbidden-energy region, below the Fermi level
$E_F=0$ eV, highlighted in green.}
\end{figure}

\subsection{GNR---base-pair---GNR junctions}
The presence of metallic electrodes in the molecular junction plays
an important role in modifying the energy levels of the
base pair through hybridization; it may result in degradation
or even loss of the rectifying mechanism. Because of the low density
of states at $E_F$, carbon-based electrodes have been proposed
as highly sensitive detectors with molecular resolution suitable for
rapid DNA sequencing \cite{Min2011} and as non-intrusive
interconnecting wires for electronics \cite{Agapito2010-1}. However,
low density of states at $E_F$ also indicates electrodes with rapidly decaying
wavefunctions and therefore small variations to the adsorption geometry of the base pair, while translocating through the nanogap, can induce large fluctuations of
the tunneling current; nonetheless, nonlinear techniques have been
proposed to overcome that difficulty \cite{Postma2010}.

We use reconstructed graphene nanoribbons (GNRs) as electrodes. The reconstructed edges are composed of heptagons and pentagons and increase the dispersion of the otherwise flat
low-energy bands; therefore, it eliminates the well-known spin-induced bandgap of
unreconstructed zigzag ribbons, thus rendering the reconstructed GNR
metallic. Reconstructed ribbons have been predicted
to be energetically favorable \cite{Koskinen2008}. We construct a nanogap of $\sim$ 5\AA\ in the reconstructed GNR, whose C edge atoms are passivated with H. The geometry of the nanogap is further relaxed and kept fixed in subsequent DNA adsorption, as seen in Figure 2a. Our transport calculations confirm that the tunneling current through the bare nanogap is negligible.

The C-G and T-A base pairs were allowed to relax on top of the
graphene nanogap. In both cases, the pyrimidine subunit (G or
A) adsorbs almost parallel above the plane of the
ribbon at $\sim$ 3\AA\, following the same Bernal's AB stacking of graphite that
minimizes $\pi-\pi$ repulsion and indicates physisorption via
electrostatic coupling, as previously reported for nucleobases
\cite{Ortmann2005,Gowtham2007}. In the absence of bias, the electronic structure of the
``infinite'' GNR---base-pair---GNR junction is obtained following the Green's function formalism, where
the left and right semi-infinite electrodes are ``mathematically
attached'' to a finite ``central region'' \cite{Agapito2007}. Figure
\ref{fig_bp} shows the central regions for the resulting C-G and T-A junctions. Because of the short screening length of the carbon electrodes, the central region includes multiple screening layers and has a length equivalent to 15 GNR unit-cells, as shown in Figure \ref{vdrop}a.

In Figure 2b we show the projected electronic density of states
(PDOS) of the infinite junction on G (red) and on C
(blue) \cite{Agapito2008}. The eigenenergies of the junction ``projected'' on the C-G base pair are shown with vertical lines. First, it is observed that the projected eigenenergies keep the same ordering
than the eigenenergies of the isolated C-G (Figure 1d). Second, the PDOS consists of a sequence of narrow peaks
that closely follow the projected eigenstates, which confirms that the
electrodes do not significantly alter the molecular properties of
the isolated base pair because of weak electrostatic coupling. Both observations further indicate that the
junction may retain the Aviram-Ratner diode behavior predicted for an isolated base pair.

The Lorentzian-like G HOMO peak, around $E_F=0$ eV, clearly derives from
\textbf{4*}, which corresponds to \textbf{4} of the isolated base pair
(asterisks denote a junction's eigenstate ``projected'' on the base pair). Therefore, according to the two-level model, the HOMO peak is expected to dominate the rectifying properties of the
junction. The PDOS on C is negligible, indicating that the HOMO peak
is spatially confined to G, analogous to the
isolated-C-G case.

\subsection{Confirmation of the Aviram-Ratner mechanism}
For most cases, it suffices to calculate the electrical response of a molecular junction only in the linear-response regime (Landauer-B\"uttiker), where the transmission channels are considered independent of the
applied bias voltage. However, the frontier transmission channels (eigenstates) in Aviram-Ratner rectifiers inherently undergo electronic changes with the applied $V$ and, therefore, it requires an explicit
recalculation of the electronic charge inside the junction at every bias-voltage point (non-linear or Keldysh regime). Following our previous developments \cite{Agapito2008,Agapito2007}, we have implemented an
in-house parallel software for the computation of the electronic transport properties under non-equilibrium conditions.

The electronic charge of the junction is represented in matrix form by the density matrix $D$. The density matrix for the central region connected to two semi-infinite electrodes and under applied bias voltage $V$ was derived in Eq. \ref{dm_app}. In a non-orthogonal basis-set it is expressed as

\begin{align}
\begin{split}
D(\mu_L,\mu_R) &= D_{eq} + D_{non-eq}\\
&=\frac{i}{2\pi}\int\limits_{-\infty}^{+\infty} d\varepsilon\,(G^{+}-G^-)f(\varepsilon - \mu_L)\\
&+\frac{1}{2\pi}\int\limits_{-\infty}^{+\infty} d\varepsilon\,[\overbrace{2\eta G^{+}SG^-}^\text{unbound} + \overbrace{G^+\Gamma_RG^-}^\text{bound}]\\
&\times[f(\varepsilon - \mu_R)-f(\varepsilon - \mu_L)],\label{ddmm}
\end{split}
\end{align}

The Hamiltonian matrix $H$ of the central region under applied $V$ is computed using a standard DFT solver (SIESTA 2.0.2 \cite{Soler2002}), modified to introduce $V$ as an external Hartree potential. Starting from an initial guess, the density matrix is calculated according to Eq. \ref{ddmm} and fed back to the DFT solver, which yields an updated Hamiltonian matrix. The process is repeated iteratively until self-consistent convergence is achieved.

\begin{equation}
D(H) \xrightleftharpoons[NEGF]{DFT solver} H(D)
\end{equation}

\begin{figure}
\includegraphics[width=.4\textwidth]{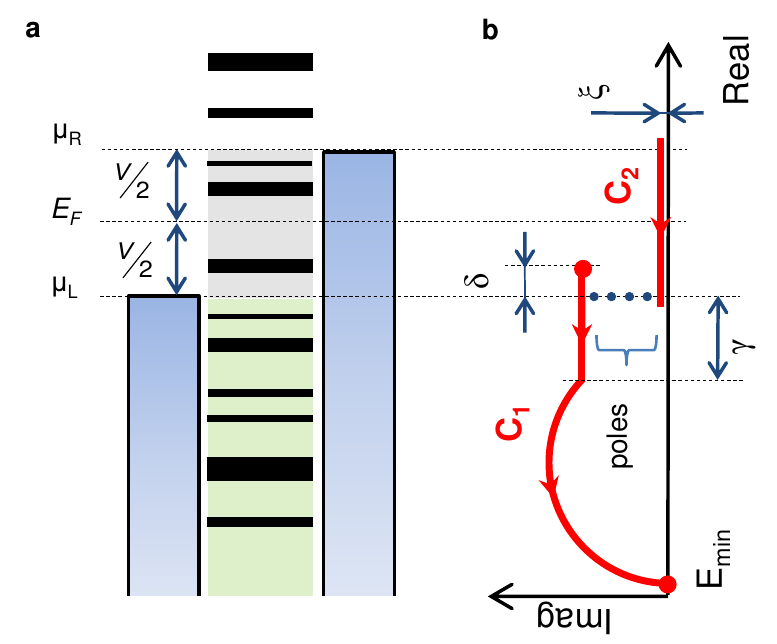}
\caption{\label{fig_complex_plane} (a) Schematic representation of the electronic states of an infinite junction under an applied external bias voltage $V$. The finite broadening of the discrete energy levels of the central region (horizontal lines) schematically accounts for the coupling to both semi-infinite electrodes. Under bias, the reservoirs of electronic states of the left and right electrodes (blue regions) are assumed to shift rigidly around the Fermi level ($E_F$). (b) The complex-energy-plane paths $C_1$ and $C_2$ used for the integration of the states in the green and gray area, respectively. The arc in $C_1$ is centered on the real-energy axis. The blue dots represent the poles of the Fermi function of the left electrode that are enclosed by $C_1$ and the real axis.}
\end{figure}

The equilibrium contribution $D_{eq}$ in Eq. \ref{ddmm} is determined by integrating the states inside the green area schematically shown in Figure \ref{fig_complex_plane}a. Because of the analytic properties of the Green's function, the integral of $\mathcal{I}(\varepsilon)= (G^{+}(\varepsilon)-G^-(\varepsilon))f(\varepsilon - \mu_L)$ can be analytically continued to the complex plane, away from the real-axis singularities. The initial real-axis path $R=[-\infty;+\infty]$ is closed with a counterclockwise arc $C$ in the upper-half complex plane extending from $+\infty$ to $-\infty$. From Cauchy's residue theorem applied to the integral of $\mathcal{I}$ along the described closed contour $R+C$, we have

\begin{equation}
\oint_{R+C} dz \mathcal{I}(z)=\int\limits_{-\infty}^{+\infty} d\varepsilon \mathcal{I}(\varepsilon) + \int_C dz \mathcal{I}(z) = 2 \pi i \sum\limits_{\nu=1}^{n_p} Res(\mathcal{I},z_{\nu}),
\end{equation}
where $Res(\mathcal{I},z_{\nu})$ is the residue of $\mathcal{I}(z)$ evaluated at the singularities $z_{\nu}$ inside the closed contour. Those singularities are assumed to be first order and to come only from the poles of the Fermi function, $z_{\nu}=\mu_L + i(2\nu+1)\pi k_BT$. The residues of the Fermi function are $Res(f,z_{\nu})=-k_BT$.

\begin{align}
\begin{split}
Res(\mathcal{I},z_{\nu}) &= \lim_{z \to z_{\nu}} (z-z_{\nu})\mathcal{I}(z)\\
&=\lim_{z \to z_{\nu}} (z-z_{\nu})f(z - \mu_L)(G^{+}(z)-G^-(z))\\
&=[\lim_{z \to z_{\nu}} (z-z_{\nu})f(z - \mu_L)](G^{+}(z_{\nu})-G^-(z_{\nu}))\\
&=Res(f,z_{\nu})(G^{+}(z_{\nu})-G^-(z_{\nu}))\\
&=-k_BT(G^{+}(z_{\nu})-G^-(z_{\nu})),\label{residue}
\end{split}
\end{align}

In practice the counterclockwise path $C$ is approximated to $C=C_1$ in Figure \ref{fig_complex_plane}b. The equilibrium contribution $D_{eq}$ in Eq. \ref{ddmm} reduces to

\begin{equation}
D_{eq}= - \frac{i}{2\pi}\int_{C_1} dz \mathcal{I}(z) + k_BT\sum\limits_{\nu=1}^{n_p} [G^+(z_{\nu})-G^-(z_{\nu})].
\end{equation}
The path $C_1$ is defined by $\gamma=20k_BT$, $\delta=10k_BT$, $\lambda=2\pi n_p k_BT$, and $E_{min}=\mu_L-30$ eV, where $n_p=5$ is the number of Fermi-function poles considered, $k_B$ is the Boltzmann constant, and
$T=300K$. The integrals along the line and arc of $C_1$ are performed using a 10-point trapezoidal integration and a 85-point Gaussian-Legendre quadrature, respectively. The infinitesimal $\eta = 10^{-8}$ eV.
The non-equilibrium contribution $D_{non-eq}$ in Eq. \ref{ddmm} involves an integration along the path $C_2$, infinitesimally away from the real-energy axis, from $(\mu_R+7k_BT)+i\xi$ to $(\mu_L-7k_BT)+i\xi$, with $\xi=10^{-8}$ eV.

Having determined a converged electron density, $D^{convg}$, for the biased junction, its electronic properties  such as PDOS, transmission-probability function ($TF(\varepsilon,V)$), and current-voltage characteristics ($I-V$) can be determined from $H(D^{convg})$ and the well-known linear-response transport equations.
\begin{equation}
TF(\varepsilon,V)=Tr(\Gamma_LG^{+}\Gamma_RG^{-}),
\end{equation}
and
\begin{equation}
I(V)=\frac{n_{\sigma}e}{h}\int\limits_{-\infty}^{+\infty} d\varepsilon\,TF(\varepsilon,V)[f(\varepsilon - \mu_R)-f(\varepsilon - \mu_L)]
\end{equation}
where $n_{\sigma}=2$ accounts for the spin degeneracy while all
remaining terms are defined in the Computational Methods section.

\begin{figure}
\includegraphics[width=.4\textwidth]{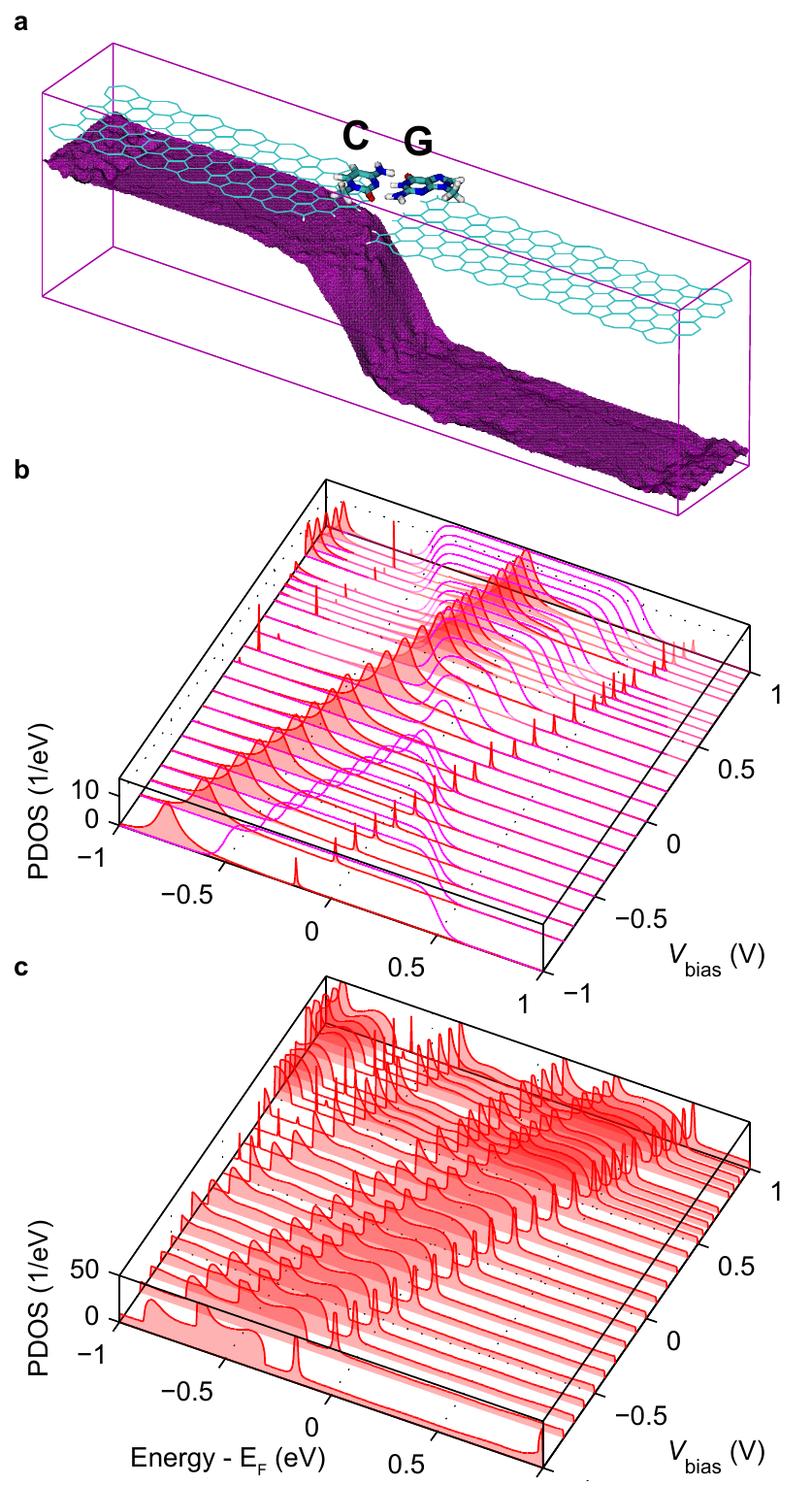}
\caption{\label{vdrop} (a) Bias-voltage drop along the central region of the C-G junction. The left and right leads in our model contains 6 and 7 GNR unit-cells, respectively. The applied bias $V=-0.5$ V drops entirely around the nanogap and not along the electrodes. (b) Bias evolution of the density of states projected on Guanine (red). The broad peak, which shifts with bias, corresponds to Guanine's HOMO. Fermi windows (magenta) at every bias voltage are shown as reference. (c) Bias-voltage evolution of the density of states projected on the right electrode. }
\end{figure}

Figure \ref{vdrop}a shows the drop of the electrostatic potential for $V =
-0.5$ V along the C-G junction. Note that the voltage drop occurs
mainly inside the nanogap, confirming that the external
potential is well screened at the boundaries of the central region. Moreover, as seen in Figure \ref{vdrop}c, the
electronic states of the right electrode shift in energy with applied voltage at a rate of $\sim 0.44$ eV/V, which is only slightly below the rigid-shift rate of $0.5$ eV/V and thus corroborates negligible voltage drop inside the right lead.
Because of electrostatic coupling to the electrodes, both subunits of the base pair follow the local electrostatic
potential of the GNR on which they are adsorbed. Figure \ref{vdrop}b
shows the G HOMO peak shifting in concert with applied
bias, although at a slightly lower rate ($\sim 0.36$ eV/V) than the right electrode. The rate mismatch indicates a small voltage drop building up at the interface between G and the tip of the electrode.

According to the two-level model, the onset of the forward current for the C-G junction is
expected at $V_F = \frac{1}{e}\Delta_{\mathbf{4*}} \sim 0.38$ V as
the HOMO peak enters the tunneling energy window. Accordingly, the transport calculations find the onset at $\sim 0.35$ V, as seen in the blue curve in Figure \ref{fig_current}b. Under reverse bias, conversely, the HOMO
peak shifts away from the tunneling window and consequently, since level \textbf{1*} does not participate, the resulting current is low and due only to nonresonant ``leakage'' tunneling (see
later discussion on reverse-bias leakage current). As a result, the $I-V$ curve of the C-G junction
adopts the rectifying shape for a diode. It is important to emphasize
that this $I-V$ feature can be used to discern whether a C-G or a
G-C base pair is translocating in the GNR nanogap. Similar
differentiation has been reported in base pairs chemically adsorbed
on gold electrodes \cite{Jauregui2009}.

\subsection{Behavior of the T-A junction}
Analogous to C-G and according to the two-level model, the T-A junction should also exhibit a rectifying $I-V$ behavior
dominated by the HOMO level \textbf{4*} of A.
However, for the T-A case, \textbf{4*} is well below the Fermi energy in Figure 2d
and hence could only be reached at rather high biases ($V_F =
\frac{1}{e}\Delta_{\mathbf{4*}} \sim 0.92$ V). Accordingly, the electrical conductance of the T-A junction is found to be much lower than C-G for the forward-bias points shown in Figure \ref{fig_current}, which confirms that the Adenine's resonant level \textbf{4*} is not contributing to conduction.

\subsubsection{Leakage current for reverse and low biases}
\begin{figure}
\includegraphics[width=.4\textwidth]{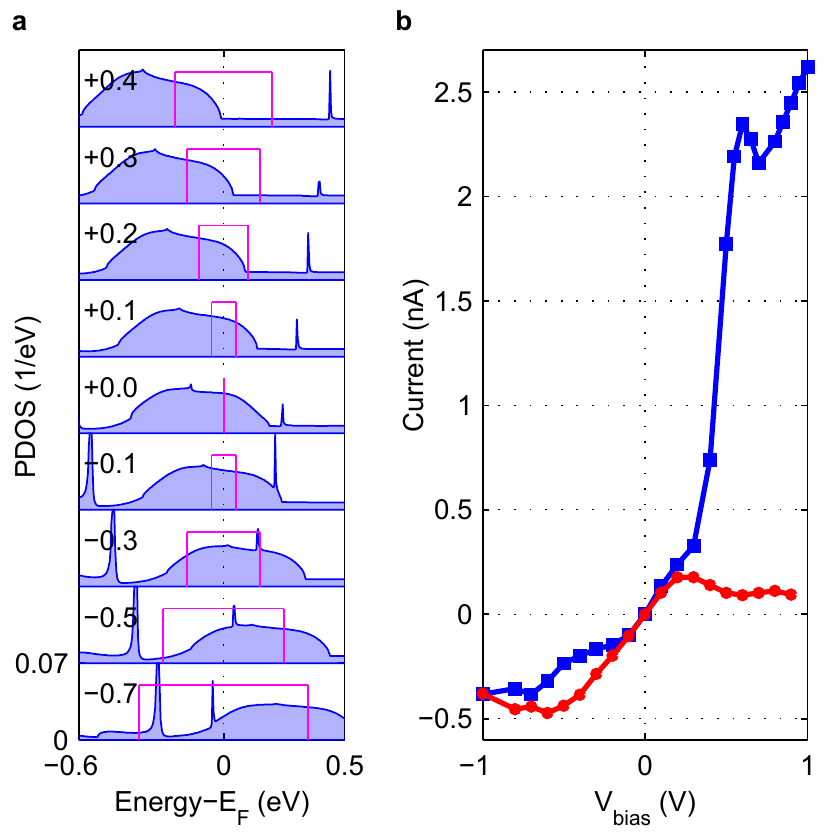}
\caption{\label{fig_current} (a) Evolution of the density of states projected on Thymine for various bias. (b) Current-Voltage ($I-V$) curves for the C-G (blue) and T-A (red) junctions.}
\end{figure}

The reverse-bias current for the T-A is small and comparable to C-G.  Such ``leakage'' current arises from double non-resonant tunneling through both subunits and is analogous to
Bardeen's tunneling theory for a scanning tunneling microscope tip
(Adenine) on top of a surface (Thymine) \cite{Tersoff1983}. The leakage
tunneling in the T-A junction is determined by the number of
electronic states available on both T and A.  Moreover,
the number of states on T is $\sim2$ two orders of magnitude
lower than those on A; thus, T constitutes the
limiting factor in the leakage-tunneling process.
Bardeen's tunneling mechanism offers a simple explanation for the sigmoidal shape of the T-A current based only on the mesa-like shape of the zero-bias T PDOS . At low energies ($|E-E_F|
\lesssim 0.1$ eV), the mesa-like T PDOS is roughly constant (Figure
\ref{fig_current}a) which , upon integration, yields the linear
(ohmic) region seen in Figure \ref{fig_current}b-red for $|V|
\lesssim 0.2$ V. Moreover, in contrast to A, the T PDOS are spatially localized on the left side of the nanogap and hence shifts towards lower energies under applied bias $V$, as seen in Figure 5a.
The number of states (area within the Fermi energy window)
progressively increases with bias  (linear region), until it
saturates at $V\sim +0.3$ V and $-0.7$ V, as seen in
Figure \ref{fig_current}a, marking the forward- and reverse-bias
saturation points of the $I-V$ red curve.

As mentioned before, for $V > V_F \sim 0.92$ V, one may expect larger resonant-tunneling currents mediated by level \textbf{4*} of A. However, there are two mechanisms that may prevent such trend:

\subsubsection{Inhibited hybridization of the HOMO level}
\begin{figure}
\includegraphics[width=.4\textwidth]{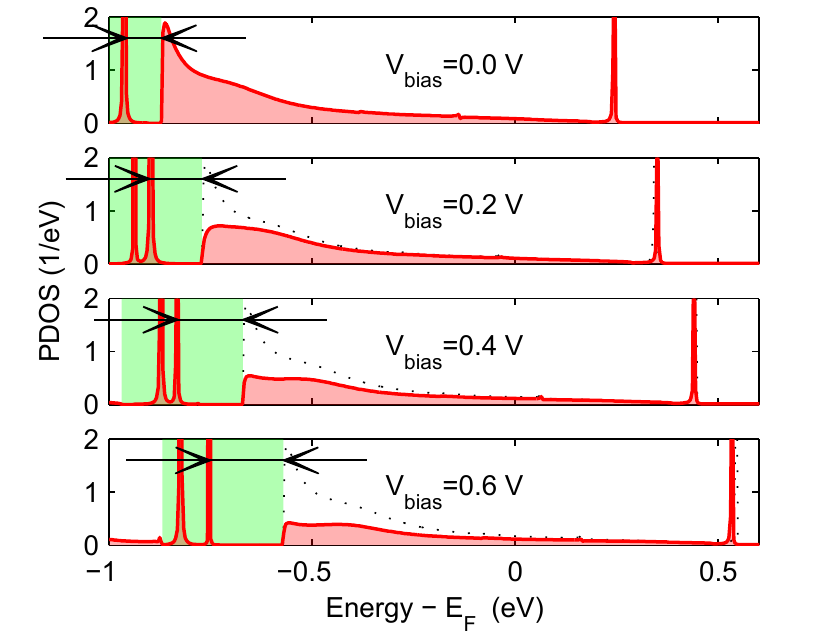}
\caption{\label{fig_a_pdos} Evolution of the truncated Adenine HOMO peak and the peak corresponding to \textbf{4*}, both shown in Figure \ref{fig_bp}d, under an applied bias voltage $V$. The arrows show the progressive energy mismatch (0.09 eV, 0.13 eV, 0.16 eV, and 0.18 eV) between both peaks.
For visual aid, the forbidden energy region of the right electrode (green) and the zero-bias PDOS (dotted lines) are shown in each panel under a rigid energy shift of $V/2$}.
\end{figure}

In contrast to G, the level \textbf{4*} of A does not fully hybridize with the states of the right electrode, since \textbf{4*} falls inside the forbidden energy region of the GNR (green highlighted area in Figure 2f). The lack of hybridization between \textbf{4*} and the right electrode hinders resonant tunneling and dramatically reduces the electrical current through the junction. Nonetheless, because of the close proximity of \textbf{4*} to the GNR states above the forbidden region (i.e. distance to the upper boundary of the forbidden region), partial hybridization is still possible at those energies. The partial or inhibited hybridization yields a truncated Lorentian-like HOMO peak in Figure \ref{fig_bp}a, which is centered at \textbf{4*}. Accounting
for a much weaker hybridization, the truncated peak in Figure 2d is $\sim 8$ times smaller than the HOMO peak in Figure 2b. Finally, a small surge in current may still be expected as the truncated peak enters the Fermi tunneling window, at large biases, analogous to the C-G HOMO peak in Figure \ref{vdrop}b; however, this trend is further prevented because the truncated peak progressively decays with increasing bias voltage, as seen in Figure \ref{fig_a_pdos}.

\subsubsection{Voltage drop at the Adenine/electrode interface}
In order to elucidate the origin of the anomalous behavior of the
A HOMO peak, we followed the evolution of the peak
and the level \textbf{4*} with bias voltage.  As seen in Figure 6, both \textbf{4*} and the HOMO peak shift toward higher
energies, but at different rates. The sharp peak corresponding to \textbf{4*} progressively lags in relation to the green forbidden region. This indicates a small voltage drop building up at the interface between A and the tip of the right electrode. With increasing bias, \textbf{4*} lags
towards the lower boundary of the forbidden region. The increasing mismatch between \textbf{4*} and the allowed states (denoted by the growing arrows in the figure) further prevents hybridization resulting in the observed decay of the A HOMO peak. Consequently, the expected surge of current at $\sim V_F$ is considerably averted.

It is important to emphasize that the unexpected departure of the T-A junction from the Aviram-Ratner behavior is solely due to the fortuitous location of Adenine's level \textbf{4*} within the forbidden energy bandgap. The bandgap itself is an artifact of the strong quantum confinement due to the ultra narrow  graphene electrodes (width of $\sim$ 1.2 nm) employed for computational tractability.  The quantum confinement in wider ribbons is weaker and the dispersion of the low-energy bands increases, effectively eliminating such bandgap \cite{Koskinen2008}.

\section{Conclusions}
We have demonstrated that in the presence of GNR electrodes, Watson-Crick DNA base pairs still act as biological Aviram-Ratner rectifiers because of the spatial separation and weak bonding between the two nucleobases.

The C-G pair junction shows a noticeable rectifying behavior with highly asymmetric $I-V$ curve, in close agreement with the Aviram-Ratner mechanism for molecular rectifiers. We propose that the $I-V$ asymmetry can be used as an electronic handle to discern C-G from G-C in the context of DNA base-pair sequencing through transverse electrical measurements using graphene nanopores.

In sharp contrast, we find that the T-A junction exhibits a rather symmetric $I-V$ with sigmoidal shape, at least for moderate voltages $|V| \lesssim 0.9$ V.
The electrical characteristics of both C-G and T-A junctions are dominated by the HOMO level of the purine nucleobase (either Guanine or Thymine). Because of the markedly
different relative positions of  their HOMO peaks with respect to $E_F$, the C-G junction exhibits larger current than the T-A junction for positive bias. For reverse bias, however,
both junctions show ``leakage'' currents of comparable magnitude.

\acknowledgements
This work was supported  by NIH Grant Nos. 1SC3GM084838-02 and
3SC3GM084838-02S1. CW and JG have been supported by NSF PREM Grant
No. DMR-0611562

\end{document}